\documentclass[epj,referee]{svjour}
\usepackage{graphics}
\begin{document}
\title{Correlation between magnetostriction and polarization in orthorhombic manganites}

\author{Radulov Iliya\inst{1,2} \and Lovchinov Vassil \inst{1}
\and Dimitrov Dimitar \inst{1}\and Nizhankovskii Viktor\inst{2}
}                     % Do not remove
\institute{"George Nadjakov" Institute of Solid State Physics;
Bulgarian Academy of Sciences; 72 "Tzarigradsko Chaussee" Blvd.;
1784 Sofia; Bulgaria Fax:+35929753632; Phone:+35929756265 \and
International Laboratory of High Magnetic Fields and Low
Temperatures; 95 Gajowicka Str.; 53421 Wroclaw; Poland
Fax:+48713610681; Phone:+48713612721 }
\date{Received: date / Revised version: date}
\abstract{Recently we have reported the observation of colossal
magnetostriction effect in HoMn$_{2}$O$_{5}$ single crystals.
Besides we have made the supposition for possible correlation
between the peculiarities, observed in the field depended
polarization measurements, and the colossal magnetostriction
effect at a 4.2 K temperature. In this article we present our
results received by polarization and magnetostriction measurements
on HoMn$_{2}$O$_{5}$ and  TbMn$_{2}$O$_{5}$ single crystals and
the strong correlation between magnetostriction and polarization
phase transition for these two compounds. The origin of this
correlation is discussed.
   \PACS{
      {75.80.+q}{Magnetomechanical and magnetoelectric effects, magnetostriction}   \and
      {75.47.Lx}{Manganites}
      {75.30.Gw}{Magnetic anisotropy}
     }
} %end of abstract
\maketitle
\section{Introduction}
\label{intro}

The study of materials which show interplay between magnetism and
ferroelectricity began in the 1960s \cite{RefSmoa,RefSmob}.
Recently a number of diverse physical phenomena (giant
magnetoresistance, giant magnetocapacitance, colossal
magnetostriction etc) in multiferroic materials were discovered.
This revival of interest in magnetoelectric materials led to the
discovery of new class multiferroic materials, in which the
magnetic order is incommensurate (IC) with the lattice period. Due
to their interesting physical properties, these compounds are
promising candidates for further practical applications.
Surprisingly, to this group of multiferroic materials belong
compounds with very various crystallographic structures like
ReMnO$_{3}$ hexagonal manganites \cite{RefFieb,RefGoto,RefArim},
ReMn$_{2}$O$_{5}$ orthorhombic manganites \cite{RefChap,RefBlak},
Ni$_{3}$V$_{2}$O$_{8}$ compounds whit Kagom\'{e}-staircase
structure \cite{RefLawe} and
Ba$_{0.5}$Sr$_{1.5}$Zn$_{2}$Fe$_{12}$O$_{22}$ hexagonal
compounds\cite{RefKimu}. A common and essential feature of these
compounds is that the frustrations in the magnetic interactions
result in non-collinear spin orderings. Generally, certain types
of magnetic order can lower the symmetry of the system to one of
the polar groups, which allows ferroelectricity. According to the
recent experimental results helical magnetic structures are the
most likely candidates to host ferroelectricity. In addition,
X-ray diffraction studies in a number of the above materials have
revealed that the modulated magnetic structure is accompanied by
structural modulation. It is, therefore, a natural assumption that
lattice displacements actively participate in the formation of the
ferroelectric (FE)state as well the FE displacements. Owing to
their smallness they have not been measured directly yet. This
calls for theoretical microscopic models providing a mechanism by
which the FE lattice displacements are induced and coupled to the
IC magnetic structure. In this paper we have try to give a
theoretical explanation for our experimental results.

\section{Samples and experiment}
\label{SandE} Single crystals of HoMn$_{2}$O$_{5}$  and
TbMn$_{2}$O$_{5}$  were grown as described elsewhere
\cite{RefRada}. The samples were characterized and oriented by
X-rays diffraction. The magnetization measurements were realized
whit Foner-type magnetometer on a frequency 3.6 Hz. Cubical
samples with typical dimensions data 1.2 x 1.4 x 1.5 mm$^{3}$ and
weights 9.8 -13.5 mg are used. For our dielectric constant
measurements thin rectangular specimens of single domain crystals
with typical area 3 - 4 mm$^{2}$, thickness 0.3 mm and weight 7.4
- 9.2 mg were used. The dielectric constant measurements were
conducted on high precision capacitance bridge AH 2550A in fields
0 - 14 T and temperatures 4.2 - 300 K. Samples polarization at
fixed H and fixed T were measured using a Keithly 617
electrometer. The magnetostriction (MS) data were obtained by use
of high precision capacitance dilatometer at different
temperatures below 100 K in fields up to ±14 T.

\section{Results and Discussion}
\label{RandD}

\begin{figure}
\resizebox{0.45\textwidth}{!}{%
  \includegraphics{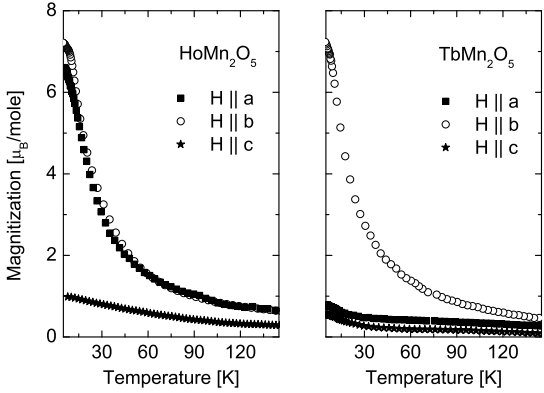}
}
 \caption{Temperature dependencies of magnetization for HoMn$_{2}$O$_{5}$
 and TbMn$_{2}$O$_{5}$ single crystals along the three principal crystallographic
 directions as following along axis \textbf{a} - square \textbf{b} - circle \textbf{c} - star
}
\label{fig:1}
\end{figure}

At room temperature ReMn$_{2}$O$_{5}$ single crystals have space
group P$_{bam}$. The structure consists of edge-sharing
Mn$^{4+}$O$_{6}$ octahedra, forming chains along the c axis,
crosslinked via Mn$^{3+}$O$_{5}$ pyramidal units. Magnetization
data for HoMn$_{2}$O$_{5}$ and TbMn$_{2}$O$_{5}$ single crystals
have been acquired as a temperature dependence in range 4.2 - 120
K and as magnetic field dependence in range 0 - 14 T. Saturation
of the sample magnetization in fields up to 14 T was not observed.
Typical temperature dependence curves for HoMn$_{2}$O$_{5}$ and
TbMn$_{2}$O$_{5}$ along the three principal crystallographic
directions are shown respectively in Fig.~\ref{fig:1} left block
and Fig.~\ref{fig:1} right block. In both compounds a significant
magnetic anisotropy is presented. The values of the total magnetic
moment of the compounds, derived from our measurements, are 17.4
$\mu_{B}$ and 14.6 $\mu_{B}$ for HoMn$_{2}$O$_{5}$ and
TbMn$_{2}$O$_{5}$ respectively, which are in a good agreement with
the expected ones.

It is characteristic for all REMn$_{2}$O$_{5}$ compounds that the
various magnetic phase changes are reflected in sharp and distinct
anomalies of the dielectric constant, as shown for
HoMn$_{2}$O$_{5}$ and TbMn$_{2}$O$_{5}$ in Fig.2.
\begin{figure}
\resizebox{0.45\textwidth}{!}{%
  \includegraphics{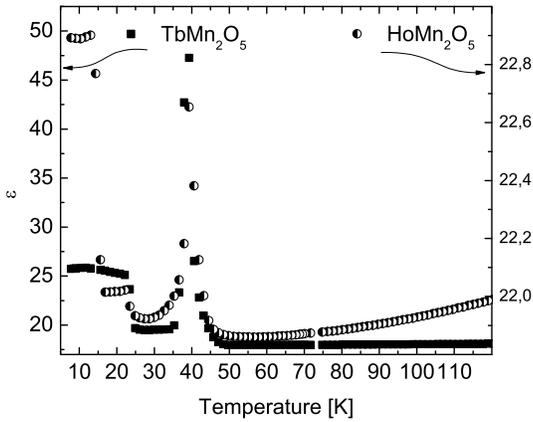}
}
 \caption{Temperature dependency of dielectric constant for TbMn$_{2}$O$_{5}$ (square)
 and HoMn$_{2}$O$_{5}$ (circle) single crystals at 1 kHz along the \textbf{b} axis
 measured by sample cooling the in presence of 1 T magnetic field.
 }
\label{fig:2}
\end{figure}
This is a clear indication for strong magneto-electric coupling
due to large spin-lattice interactions. Long-range
antiferromagnetic (AFM) ordering of the Mn$^{3+}$/Mn$^{4+}$ spins
occur at T$_{N}$ = 43 K. This transition into a high N\'{e}el
temperature phase is the common features for all
REMn$_{2}$O$_{5}$. Subsequently the FE transition takes place at
T$_{C}$ slightly below T$_{N}$ (T$_{C}$ = 39 K for
HoMn$_{2}$O$_{5}$ and 38 K for TbMn$_{2}$O$_{5}$). The pure
ferroelectric lock-in transition, observed at 39 K is not
influenced by magnetic fields.With further temperature decreasing,
at T'$_{N}$ (~ 22 K for HoMn$_{2}$O$_{5}$ and ~ 24 K for
TbMn$_{2}$O$_{5}$)  another magnetic transition takes place, at
which commensurate AFM ordering becomes low temperature
incommensurate. This transition is accompanied by a significant
decrease of the FE polarization and is often referred as a second
FE phase transition. Under T'$_{N}$  the spin wave vector remains
unchanged and the transition involves an increase in the ordered
moments of the Mn$^{4+}$/Mn$^{3+}$ sublattice. Below 19 K a phase
transition to canted AFM (CAFM I) takes place in HoMn$_{2}$O$_{5}$
and TbMn$_{2}$O$_{5}$ single crystals. It was found from our
measurements \cite{RefRada,RefRadb} that second CAFM - type
ordering (CAFM II) of RE ions occurs at T$_{N}$(Ho) below 11 K.
Measurements at low magnetic fields show peculiarities in the
dielectric constant for both compounds around 11 K. The last two
transitions, at T$_{N}$(Ho,Tb) and T'$_{N}$, change significantly
their shape and place, depending on the intensity of the applied
magnetic field. This is a clear indication for their magnetic
origin. It has been assumed that the long-range magnetic ordering
of Mn$^{3+}$/Mn$^{4+}$ induces the FE transition via an additional
Jahn-Teller distortion of Mn$^{3+}$ ions \cite{RefKenz}. The FE
state exhibits canted antiferromagnetic displacements of the
Mn$^{3+}$ ions. These displacements lift the magnetic degeneracy
by lowering the crystal symmetry to P$_{b21m}$, thus stabilizing
the FE state via the magnetic Jahn-Teller effect.
\begin{figure}
\resizebox{0.45\textwidth}{!}{%
  \includegraphics{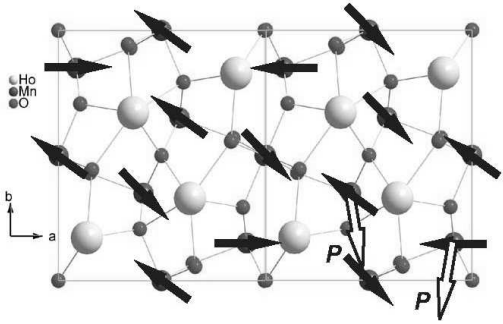}
} \caption{Spin and dipolar moment (black and white arrows)
orientation in doubled along the \textbf{a} axis ReMn$_{2}$O$_{5}$
unit cell} \label{fig:3}
\end{figure}
As shown on Fig~\ref{fig:3} The spins of two Mn$^{3+}$ per unit
cell are each frustrated with two neighboring Mn$^{4+}$ with the
same spin direction. Reducing this frustration by moving the
Mn$^{3+}$ away from the Mn$^{4+}$ generates a dipolar moment
\textbf{P} \cite{RefRadb} between the Mn$^{3+}$ and the
surrounding oxygen ions. The \textbf{P$_\textbf{a}$} components of
this dipolar moments cancel out while the
\textbf{P$_\textbf{b}$}-components add up to the macroscopic
polarization and ferroelectricity along the \textbf{b} axis. The
proposed displacement lowers the symmetry to the space group
P$_{b21m}$. The AFM modulation along the \textbf{a} axis with
\textbf{q}$_{x}$=0.5 leads to the frustration and displacement of
both Mn$^{3+}$ and the net polarization along the \textbf{b} axis.
In the temperature range 4.2 $-$ 43 K a consistency of
magneto-elasto-electric phase transitions was observed. It was
pointed out that all the ferroelectric phases are strongly tied to
the antiferromagnetic Mn$^{3+}$/Mn$^{4+}$ spin structure, with the
latter being dominated by the f-d exchange interaction
\cite{RefFieb}.

\begin{figure}
\resizebox{0.45\textwidth}{!}{%
  \includegraphics{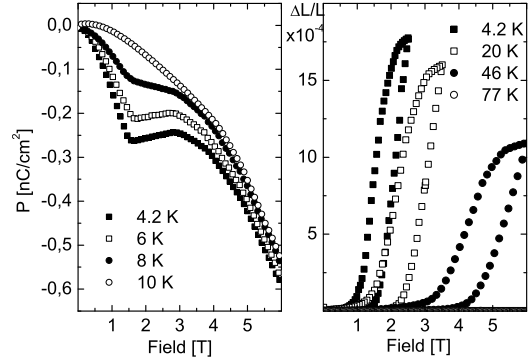}
} \caption{Polarization (left block) and magnetostriction (right
block) field dependencies measured along the \textbf{b} axis of
HoMn$_{2}$O$_{5}$ single crystals } \label{fig:4}
\end{figure}
The appearance of ferroelectricity is a consequence of frustration
between NN and NNN (next-nearest neighbour) Mn$^{4+}$ in the
lattice. The frustration is lifted by Jahn-Teller distortion, and
the associated reduction of symmetry allows the formation of a
spontaneous polarization. Considering the role of magnetic
frustration to stabilize the ferroelectricity in ReMn$_{2}$O$_{5}$
there are interesting similarities to multiferroic
Ni$_{3}$V$_{2}$O$_{8}$ and TbMnO$_{3}$ \cite{RefKenz}. By other
compounds it was shown that the transition sinusoidal to helical
magnetic modulation can introduce a third order coupling giving
rise to FE order \cite{RefKenz}. On the other hand more detailed
treatment shows that the existence of a spiral magnetic structure
alone is not yet sufficient for FE: not all the spiral can lead to
it. As shown in \cite{RefMost} FE can appear if the spin rotation
axis \textbf{e} does not coincide with the wave vector of a spiral
\textbf{Q}: the polarization \textbf{P} appears only if these two
directions are different and it is proportional to the vector
product of \textbf{e} and \textbf{Q} : \textbf{P} $\sim$
\textbf{Q} x \textbf{e}. However, it is not clear yet if the
magnetic structure between T$_{C}$ and T$_{N}$ for
HoMn$_{2}$O$_{5}$ and TbMn$_{2}$O$_{5}$ is sinusoidal and the
transition into the FE phase follows the same mechanisms as in
Ni$_{3}$V$_{2}$O$_{8}$ or TbMnO$_{3}$. Furthermore, the magnetic
modulation in the FE phase of HoMn$_{2}$O$_{5}$ and
TbMn$_{2}$O$_{5}$ is commensurate whereas it is incommensurate in
Ni$_{3}$V$_{2}$O$_{8}$ or TbMnO$_{3}$.
\begin{figure}
\resizebox{0.45\textwidth}{!}{%
  \includegraphics{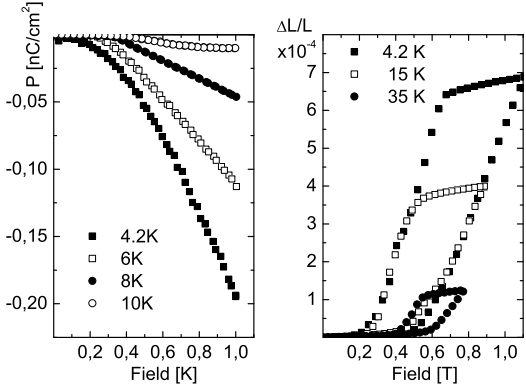}
} \caption{Polarization (left block)and magnetostriction (right
block) field dependencies measured along the \textbf{b} axis of
TbMn$_{2}$O$_{5}$ single crystals } \label{fig:5}
\end{figure}
As discussed in \cite{RefRada} the reason to observe such colossal
magnetostriction effect is the total effect of the exchange
magnetostriction of the manganese ions and the holmium single ion
magnetostriction. Holmium and Terbium in metal state showed
gigantic single-ion magnetostriction \cite{RefNiki}, which is due
to both, the strong spin-orbit coupling between orbital magnetic
moment M$_{L}$ and non-spherical charge cloud of \textbf{4f} -
electron shell (which is highly anisotropic), and strong
spin-lattice interactions. When a Ho ion is placed in the crystal
lattice the anisotropy of the \textbf{4f}-electron shell remains
practically unchanged. In external magnetic field the spin moment
M$_{S}$ changes its orientation and this leads to reorientation of
M$_{L}$. This causes a strong perturbative effect on the crystal
field (the spin - lattice interactions in the HoMn$_{2}$O$_{5}$
and TbMn$_{2}$O$_{5}$ compound are strong \cite{RefCrus}) and a
colossal magnetostriction effect appears.

The polarization and magnetostriction field dependencies of
HoMn$_{2}$O$_{5}$ and TbMn$_{2}$O$_{5}$ single crystals are shown
on Fig.~\ref{fig:4} and Fig.~\ref{fig:5}  respectively. As will
readily be observed, especially for HoMn$_{2}$O$_{5}$, the place
of the peculiarities in polarization field dependencies and the
magnetostriction phase transition at 4.2 K is nearly the same.
More detailed measurements in the temperature range 4.2 - 10 K
have corroborate our assumption for possible correlation between
the peculiarities in polarization and the magnetostriction phase
transition. Our precise magnetization measurements in this
temperature range allows to observe the absence of any
magnetization peculiarities for HoMn$_{2}$O$_{5}$ neither for
TbMn$_{2}$O$_{5}$. In our opinion, the observation of polarization
peculiarities only below 10 K is a direct consequence of the RE
(Ho,Tb) spin reorientation.

\begin{figure}
\resizebox{0.45\textwidth}{!}{%
  \includegraphics{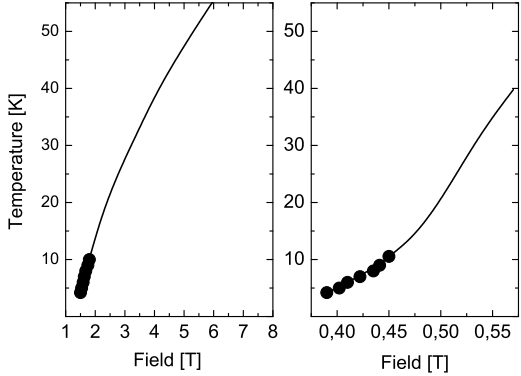}
}
\caption{Phase diagram of HoMn$_{2}$O$_{5}$ left block and
TbMn$_{2}$O$_{5}$ right block single crystals built from data
received by magnetostriction (solid line) and polarization
(circles) measurements along the \textbf{b} axis. }
\label{fig:6}
\end{figure}

The phase diagrams build from magnetostriction (line) and
polarization (circle) measurements of HoMn$_{2}$O$_{5}$ and
TbMn$_{2}$O$_{5}$ single crystals are shown on Fig.~\ref{fig:6}
left and right block respectively.

The strong magnetoelectric correlation is indicated by the
observation that ordering of the Mn spins modifies the dielectric
function, while ferroelectric ordering leaves an imprint on the
magnetic susceptibility \cite{RefFieb}. Magnetoelastic coupling
through the exchange interaction can produce improper
ferroelectricity in systems with suitable symmetry. Within the
underlying exchange interactions both symmetric and antisymmetric
(or Dzyaloshinskii-Morya) type exchange can produce
ferroelecticity \cite{RefDzya,RefMory}. The geometric magnetic
frustration among the Mn$^{3+}$/Mn$^{4+}$ spins in
REMn$_{2}$O$_{5}$ leads to a ground state degeneracy of the
magnetic states. This frustration is lifted by Jahn-Teller
distortion, and the associated reduction of symmetry allows the
formation of a spontaneous polarization. Since this polarization
is derived from canting of electric dipole moments in an
antiferroelectric arrangement, denotation as 'weak' ferroelectric
polarization may be used in analogy to the 'weak' ferromagnetic
magnetization accompanying antiferromagnetism in the presence of a
Dzyaloshinskii-Moriya interaction (DMI) \cite{RefSchm,RefHous}.
Nowadays exist two alternative scenaria about the role of the DMI,
that linearly dependent on the displacements of the oxygen ions
surrounding transition metal ions, in the magnetoelectric effect
in IC multiferroics. In the first one is suggested that the DMI
induces the polarization of the electronic orbitals, without the
involvement of the lattice degrees of freedom \cite{RefSerg}. The
alternative scenario assert, that the DMI effect is twofold: it
induces the FE lattice displacements and helps to stabilize
helical magnetic structures at low temperature \cite{RefKats}. The
distortion of perovskite lattice leads to the further-neighbor
exchange interactions and nontrivial magnetic structures, like the
helical spin structure observed in IC, with the lattice period
magnetic order, multiferroic materials. This way the key role of
the helical spin structure, induced by frustrated exchange
interactions, in producing the electric polarization and enhanced
magneto-electric coupling is shown.

\section{Conclusions}
\label{concl}

In the present article the temperature dependencies of the
magnetization and the dielectric constant, as well the field
dependencies of the polarization and magnetostriction for two
series of HoMn$_{2}$O$_{5}$ and TbMn$_{2}$O$_{5}$ monocrystals are
discussed. As is evident, by temperature lowering these
orthorhombic manganites undergo cascade phase transitions, which
complexity origin in the partially competing interactions between
Mn$^{3+}$/Mn$^{4+}$ spins, rare earth magnetic moments and the
lattice \cite{RefRadb}. We have observed also a colossal
magnetostriction effect for both HoMn$_{2}$O$_{5}$and
TbMn$_{2}$O$_{5}$ monocrystals. This effect result from the
reorientation of the spin moments in external magnetic field and
its strong perturbative effect on the crystal field. Compared to
HoMn$_{2}$O$_{5}$ monocrystals the observed magnetostriction
effect in TbMn$_{2}$O$_{5}$ monocrystals is stronger and appears
in lower magnetic fields. The same magnetostrictive behavior was
observed in pure Ho/Tb monocrystals in metal state \cite{RefNiki}.
The more detailed measurements of magnetization, polarization and
magnetostriction of HoMn$_{2}$O$_{5}$ and TbMn$_{2}$O$_{5}$
monocrystals in the temperature range 4.2 - 10 K give us the
possibility to build the phase diagrams for both monocrystals.
These diagrams clearly demonstrate the correlation between the
peculiarities in polarization and the magnetostriction phase
transition. In our opinion, the field couples to the magnetic
order resulting in field-induced spin reorientations and magnetic
phase transitions, which in turn should affect the lattice, via
the spin-lattice interaction. In this way they realize the
correlation observed between the polarization and
magnetostriction. Knowing the signature and the nature of the
magnetoeleastic effect we have essential information about the
intrinsic magnetoelastic and magnetoelectric interactions. To
elucidate the role of the rare earth ions in this processes
further investigations are necessary.

\section{Acknowledgements}
\label{Acknow} The work of I. Radulov is supported by NATO
EAP.RIG.981824.

\end{document}